\documentclass[%
reprint,
superscriptaddress,
groupedaddress,
 amsmath,amssymb,
 aps,
]{revtex4-2}

\usepackage{graphicx}
\usepackage{dcolumn}
\usepackage{bm}
\usepackage{natbib}
\usepackage{hyperref}

\begin{document}

\title{Effective circuit modelling and experimental realization of an ultra-compact self-rectifier flux pump}

\author{B.P.P. Mallett}
\email{ben.mallett@vuw.ac.nz}
\affiliation{Paihau Robinson Research Institute, Victoria University of Wellington, P.O. Box 600, Wellington, New Zealand}
\author{S. Venuturumilli}
\affiliation{Paihau Robinson Research Institute, Victoria University of Wellington, P.O. Box 600, Wellington, New Zealand}
\author{J. Geng }
\affiliation{Wuhan National High Magnetic Field Center, Huazhong University of Science and Technology, 430074 Wuhan, People’s Republic of China}
\author{J. Clarke }
\affiliation{Paihau Robinson Research Institute, Victoria University of Wellington, P.O. Box 600, Wellington, New Zealand}
\author{B. Leuw }
\affiliation{Paihau Robinson Research Institute, Victoria University of Wellington, P.O. Box 600, Wellington, New Zealand}
\author{J.H.P. Rice  }
\affiliation{Paihau Robinson Research Institute, Victoria University of Wellington, P.O. Box 600, Wellington, New Zealand}
\author{D.A. Moseley }
\affiliation{Paihau Robinson Research Institute, Victoria University of Wellington, P.O. Box 600, Wellington, New Zealand}
\author{C.W. Bumby }
\affiliation{Paihau Robinson Research Institute, Victoria University of Wellington, P.O. Box 600, Wellington, New Zealand}
\author{R.A. Badcock}
\affiliation{Paihau Robinson Research Institute, Victoria University of Wellington, P.O. Box 600, Wellington, New Zealand}


\date{\today}

\begin{abstract}
    This paper presents experimental and modelling results of an ultra-compact self-rectifier flux pump energizing a superconducting coil. The device fits inside a volume of 65x65x50~mm and generates up to 320~A dc through the coil and a peak output voltage up to 60~mV. We also develop and present a full electromagnetic effective circuit model of the flux pump and compare its predictions to the experimental results. We show that our model can reproduce accurately the charging of the load coil and that it reproduces the systematic dependence of the maximum load coil current on the input current waveform. The experiments and modelling together show also the importance of dc-flux offsets in the transformer core on the final achievable current through the coil. The miniaturization possible for this class of flux pump and their minimal heat-leak into the cryogenic environment from thermal conduction make them attractive for applications with demanding size, weight and power limitations. Our effective circuit model is a useful tool in the understanding, design and optimization of such flux pumps which will accelerate their progression from research devices to their application. 
\end{abstract}
\maketitle

\section{Introduction}

Flux pumps (FP) are superconducting devices which output a large dc current to a superconducting coil or magnet \cite{van_de_klundert_fully_1981}. Self rectifying FPs are a subset of such devices that rectify an asymmetric, alternating input current to a transformer to generate the dc current through a superconducting load \cite{geng_hts_2016, geng_maximising_2020, zhou_contactless_2020, deng_performance_2020, zhai_performance_2021, iftikhar_2022}. Rectification is achieved by driving current larger than the critical current, $I_c$, through part of the superconducting circuit (the `bridge') for part of the input waveform cycle, as originally described for high-temperature superconductors by Vysotsky \textit{et al.} \cite{vysotsky_possibility_1990}. This self-rectification method can be compared with rectifiers that employ various active switching mechanisms to vary the $I_c$ of the bridge \cite{ van_de_klundert_fully_1981, oomen_hts_2005, geng_kilo-ampere_2019, geng_wireless_2021, leuw_half-wave_2022}. Self-rectifying FPs are capable of delivering large currents at modest voltages \cite{geng_maximising_2020} without a requirement for direct electrical and thermal connection between the superconducting load and an ambient temperature power supply. Together with the absence of moving parts and relatively few components, they may be an attractive alternative to conventional power supplies, or other types of flux pumps \cite{van_de_klundert_fully_1981, coombs_superconducting_2019, wen_high_2022}, for applications with demanding size, weight and power requirements. 

Previous experimental work on this class of flux pump has focused on demonstration of principle \cite{geng_hts_2016} or maximizing the delivered dc current \cite{geng_maximising_2020} rather than its potential advantages, comprising a simple device architecture and physical compactness. This previous work successfully described and developed what we believe to be the essential operating principles of the FP \cite{geng_maximising_2020, zhai_performance_2021}. However, quantitative and accurate prediction of the performance of these FPs involves numerically solving the coupled equations describing them. Recent numerical modelling work utilizing an effective-circuit (or `lumped parameters') methodology has been limited to exploring the influence of sub-components of the FP, such as the bridge \cite{li_hts_2020} or the transformer \cite{li_impacts_2021}. Combining these subsystems into the same model has only just been reported by Zhai \textit{et al.} \cite{zhai_modeling_2022} with some explanatory success for an air-core transformer self-rectifier. What such modelling should enable however is; (i) quantitative performance prediction to facilitate model validation, and (ii) an understanding of the interplay between all subcomponents of the FP such as the input waveform, transformer and circuit impedances. Together these would allow for useful full-system design and optimization, which has been identified as an important issue to be addressed for flux pumps \cite{wen_high_2022}. The utility of such an effective-circuit approach for system-level modelling and has recently been demonstrated in other areas of superconducting power electronics \cite{perez-chavez_generic_2019, zhang_power_2016, baez-munoz_thermoelectromagnetic_2021, de_sousa_thermalelectrical_2014}.

In this work, we construct a simple prototype `ultra-compact’ self-rectifier FP and demonstrate its functionality. We then develop a system-level effective circuit model of the FP and compare its predictions to experimental results. Our first finding is that the FP does not require a superconducting secondary-winding and can deliver more than 300 A dc to a superconducting coil with up to 60~mV peak output voltage. The effect of modifying the input current waveform is explored and the implications on performance discussed.  Our second finding is that our effective circuit model matches experimental behaviour and is capable of describing the maximum load current and its dependence on the input current to the transformer as well as the magnitude and time dependence of the voltage across the load. The paper concludes with a discussion of insights from the model regarding the operation of self-rectifying flux pumps, particularly the effect of dc offsets of flux in the transformer core, and the limitations of our current model. 

\section{Methods}
\begin{figure}
    \centering
    \includegraphics[width= 6cm]{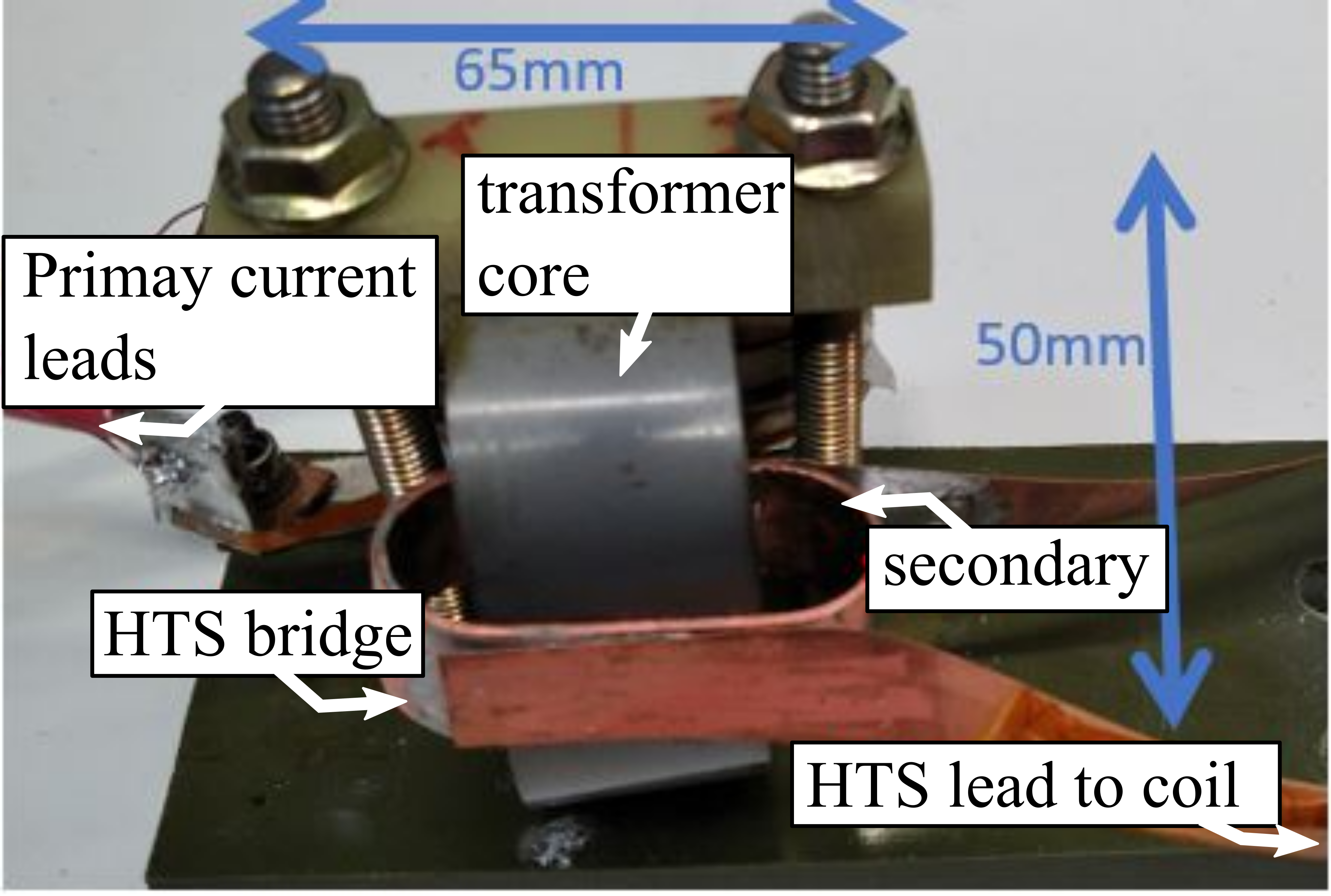}
    \includegraphics[width=7.5 cm]{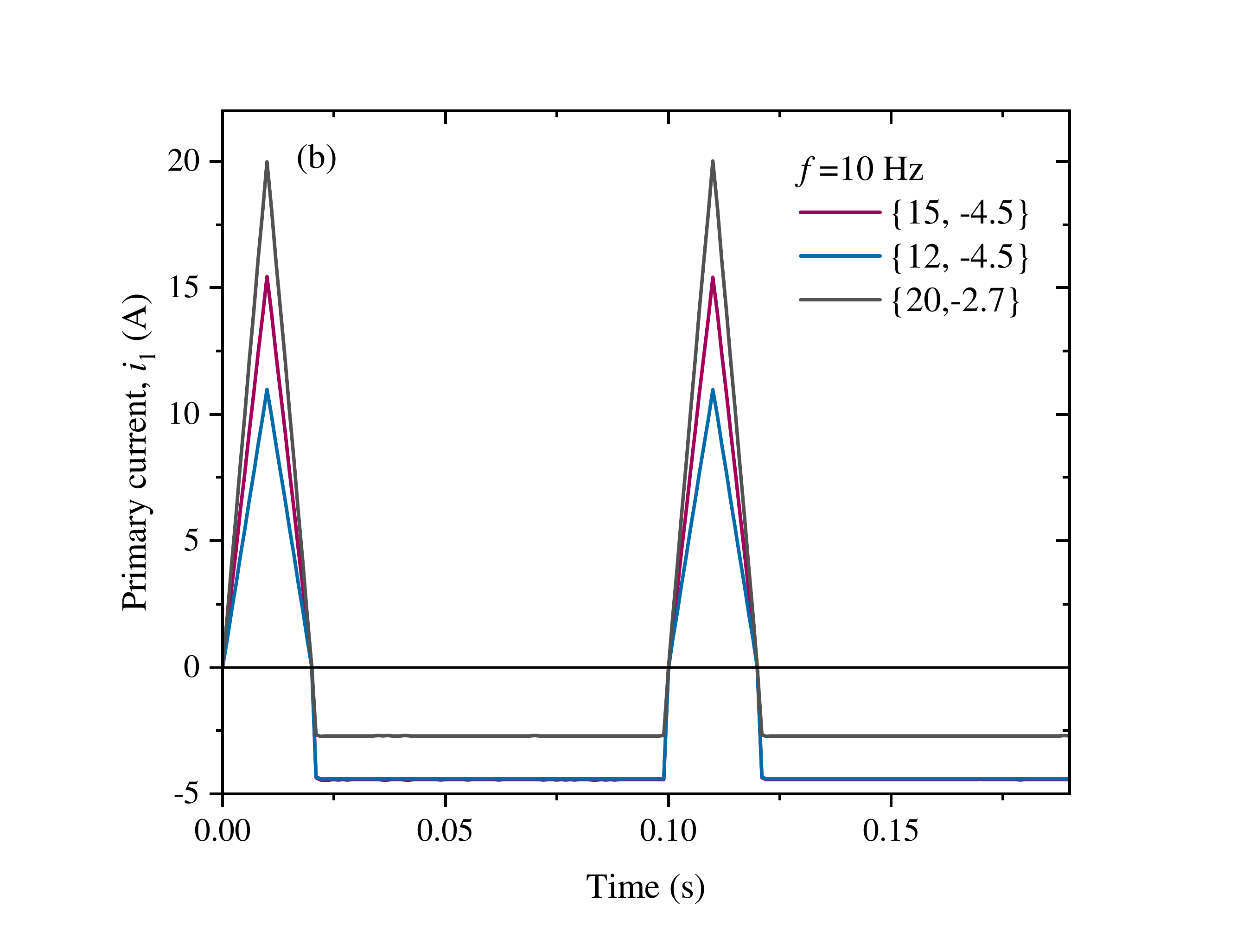}
    \caption{\label{fig:1}(a) An annotated picture of the `ultra-compact’ self-rectifier and test load coil. (b) Examples of the current waveform supplied to the primary windings of the transformer.}
    
\end{figure}

The self-rectifying FP studied in this work is shown in Fig.~\ref{fig:1}(a). It is built around an iron core made from 0.3~mm laminations with cross section of 20~mm x 20~mm and path length approximately 100~mm.  The nominal saturation field at room temperature of the soft iron is $B_{\textrm{sat}} = 1.8$~T. The primary windings are made from 45 turns of Cu tape. The secondary is made also from pure Cu and comprises a single strip, 12~mm wide, 35~mm long and 3~mm thick, bent in a U-shape around the core. Soldered to the secondary winding is a 100~mm long bridge made from commercial REBCO tape, which completes the loop around the core and results in a transformer turns ratio of $N=45$. The coated conductor tape is SuNAM product code SCN12500-210222-01; 12 mm wide, stabilized by 20~$\mu$m of Cu and has a nominal self-field $I_c$ of 500 A at 77 K with $n=35$ (see eq.~\ref{eq:1} below for the definition of $n$). A small load coil of eight turns of the same REBCO tape with a total length of $1.9$~m and an inductance of $2.5$~$\mu$H, measured at room temperature, was soldered to the bridge.

In all experiments, the current through the primary windings was supplied by a Takasago BWS 40-15 bipolar linear amplifier controlled via a National Instruments c-DAQ. Examples of the applied primary current waveform, $i_1$, are shown in Fig.~\ref{fig:1}(b), where the maximum and minimum of the current are specified by the two numbers in braces in the legend. For example, $i_1\{19.8,-4.45 \}$ indicates an input waveform with a maximum current peaking at $i_{1,max}= 19.8$~A and a current in the reverse direction of $i_{1,min}=-4.45$~A. All waveforms used in this work were driven at a frequency of 10~Hz, and all measurements were carried out in liquid nitrogen (approximately 77~K). Current through the load-coil was measured via a home built open-loop hall sensor employing a P15A sensor from Advanced Hall Systems Ltd. Current supply to the primary was measured via the voltage developed across a thermally-sunk 1~$\Omega$ shunt resistor. The voltage across the primary and load were also measured.

\begin{figure*}
\centering
	\includegraphics[width=12 cm]{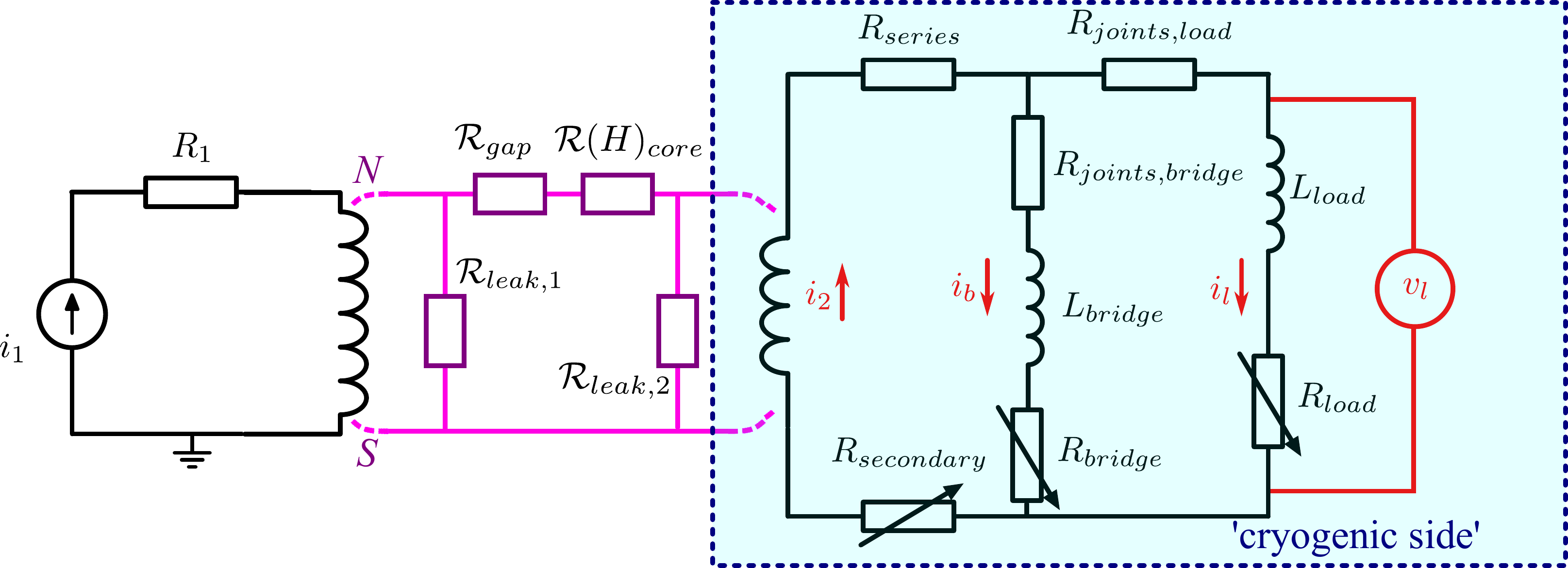}
	\caption{\label{fig:2} The electromagnetic effective circuit used to model the self-rectifier flux pump. The electric circuit is indicated in black, and coupled magnetic circuit coupled in magenta. The superconducting components are indicated as variable resistors and must be kept at cryogenic temperatures during operation. $v_l$ indicates where in the circuit the load voltage was measured in the experiment.}
\end{figure*}

The operation of the self-rectifying flux pump was modelled using the effective-circuit shown in Fig.~\ref{fig:2}. This circuit involves coupled electrical and magnetic circuits with non-linear componentry. It was implemented and solved within the Simulink package in MatLab. The current waveform input into the primary windings is specified, along with a parameterization of the circuit components.

\begin{figure}
\centering
	\includegraphics[width=0.6\linewidth]{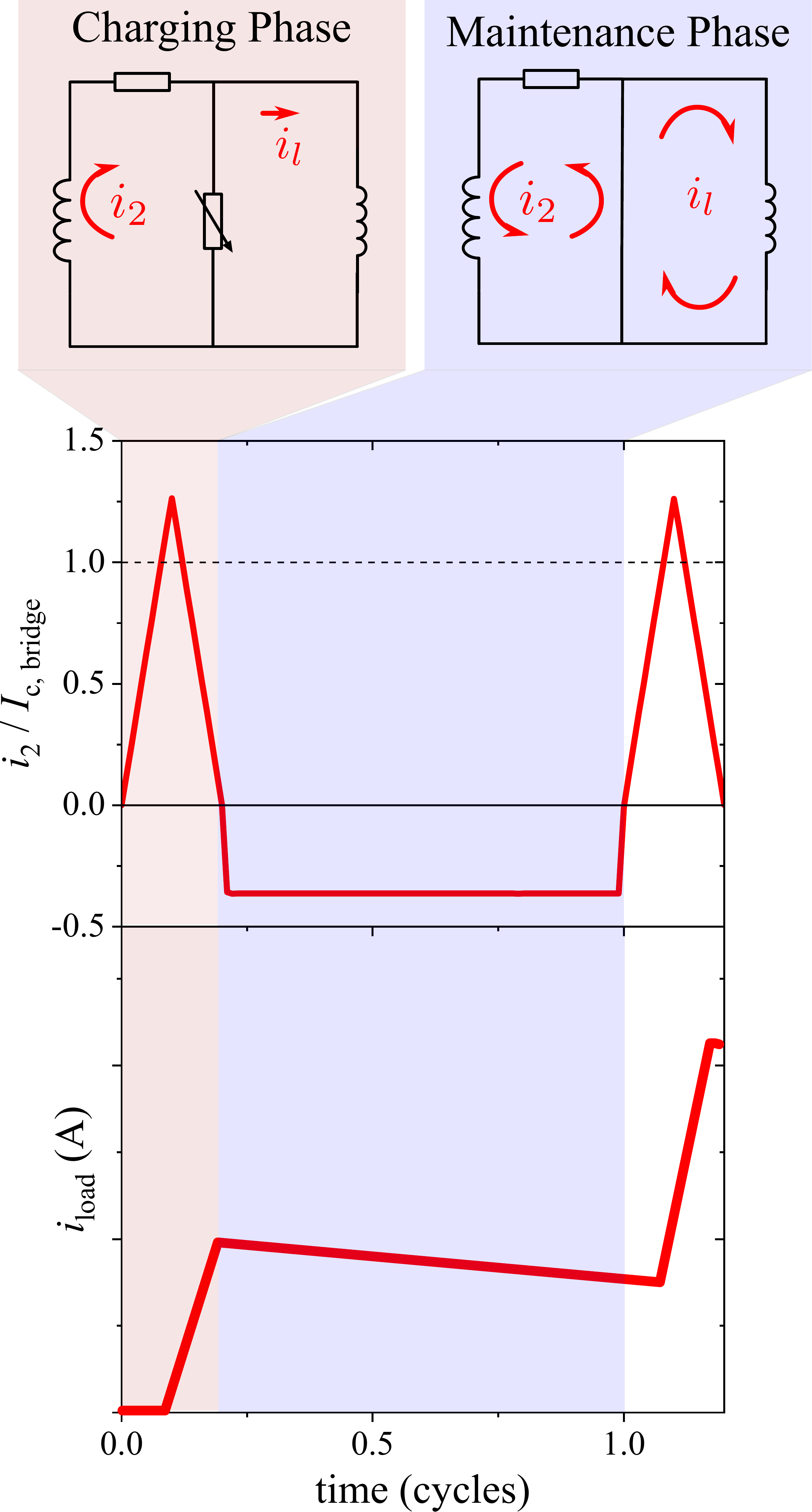}
	\caption{\label{fig:2diagram} A simplified diagram of current direction in the `charging' phase (shaded orange) and `maintenance' phase (shaded blue) of operation along with schematic . }
\end{figure}

The transformer is modelled as having a variable reluctance, $\mathcal{R}(H) = l.\mu(H)^{-1}.A^{-1}$, resulting from a magnetic material with non-linear permeability, $B(H)=\mu(H)H = B_{\textrm{sat}}\tanh(H/H_{\textrm{sat}})$, and the physical dimensions of the core length, $l$, and cross sectional area, $A$. In addition, we include in the model a leakage reluctance representing alternate paths outside of the transformer core that magnetic flux may take. This leakage reluctance is significant only as the flux in the transformer core approaches saturation. Ac-loss and hysteresis effects in the transformer core are not presently included in the model.

The superconducting components are treated as circuit elements with a resistance, $R_{\textrm{SC}}$, that is a non-linear function of the current, $I$ \cite{brandt_susceptibility_1997};   
\begin{equation}
R_{\textrm{SC}} =   \frac{E_0.l}{I_c} \left( \frac{I}{I_c} \right)^{n-1}      
\label{eq:1}
\end{equation}

\noindent Where $E_0= 1$~$\mu$V.cm$^{-1}$ is the customary electric field criterion at which the critical current density of a superconductor, $J_c$, is defined. $n$ is obtained from fitting characteristic experimental data \cite{wimbush_public_2017} and captures the steepness of the resistivity increase as $I$ exceeds $I_c$. $l$ is the length of the REBCO tape. We use Cu-stabilized tape in which the Cu represents a parallel current path. This is included as a parallel resistor in the model, such that the full expression for the resistance of a component of coated conductor (CC) tape in the model is:

\begin{equation}\label{eq:2}
R_{\textrm{CC}}^{-1}  = R_{\textrm{Cu}}^{-1} + R_{\textrm{SC}}^{-1} 
\end{equation}

\noindent where

\begin{equation}
R_{\textrm{Cu}} =  \frac{ \rho_{\textrm{Cu}}.l} { \tau_{\textrm{Cu}}.w } 
\end{equation}

In our case, we take $\tau_{\textrm{Cu}} = 20$~$\mu$m (the topmost stabilizing layer adjacent the superconductor only, as there is only a small cross-section of Cu connecting the bottom Cu layer to the superconductor), the width of the coated conductor $w=12$~mm. For high purity Cu, $\rho_{\textrm{Cu}}= 1.9\times10^{-9}$~$\Omega$.m \cite{matula_electrical_1979}. 
A look-up table is created during the initialization of the model for each coated conductor element that relates its resistance to the current (and potentially additional variables) through the element. A fuller description of the model, for the more general case of a $J_c(B)$ switched half-wave rectifier, is described elsewhere \cite{venuturumilli_submitted_2022}.

\section{Results}

\begin{figure}[htb]
\centering
	\includegraphics[width=8 cm]{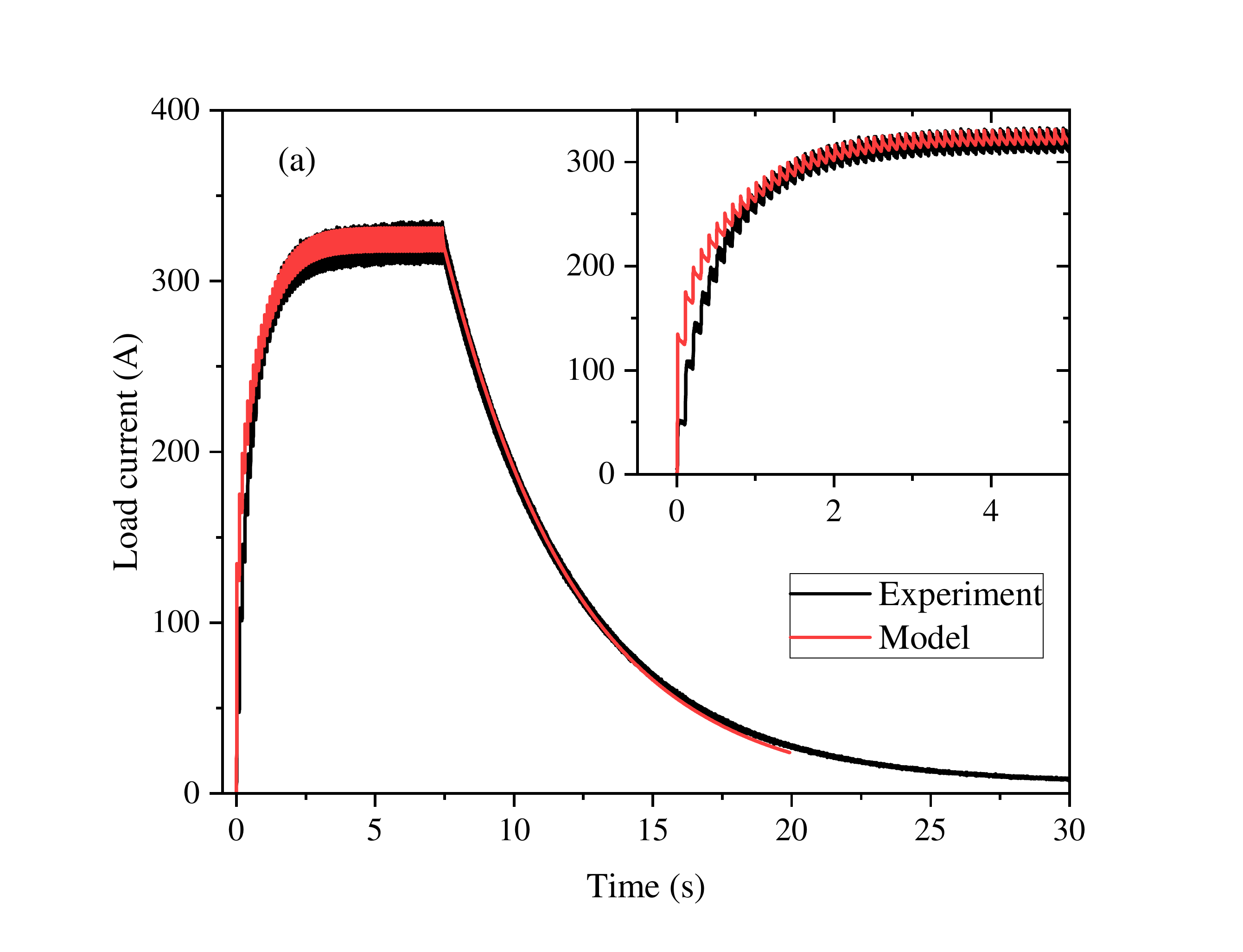}
	\includegraphics[width=8 cm]{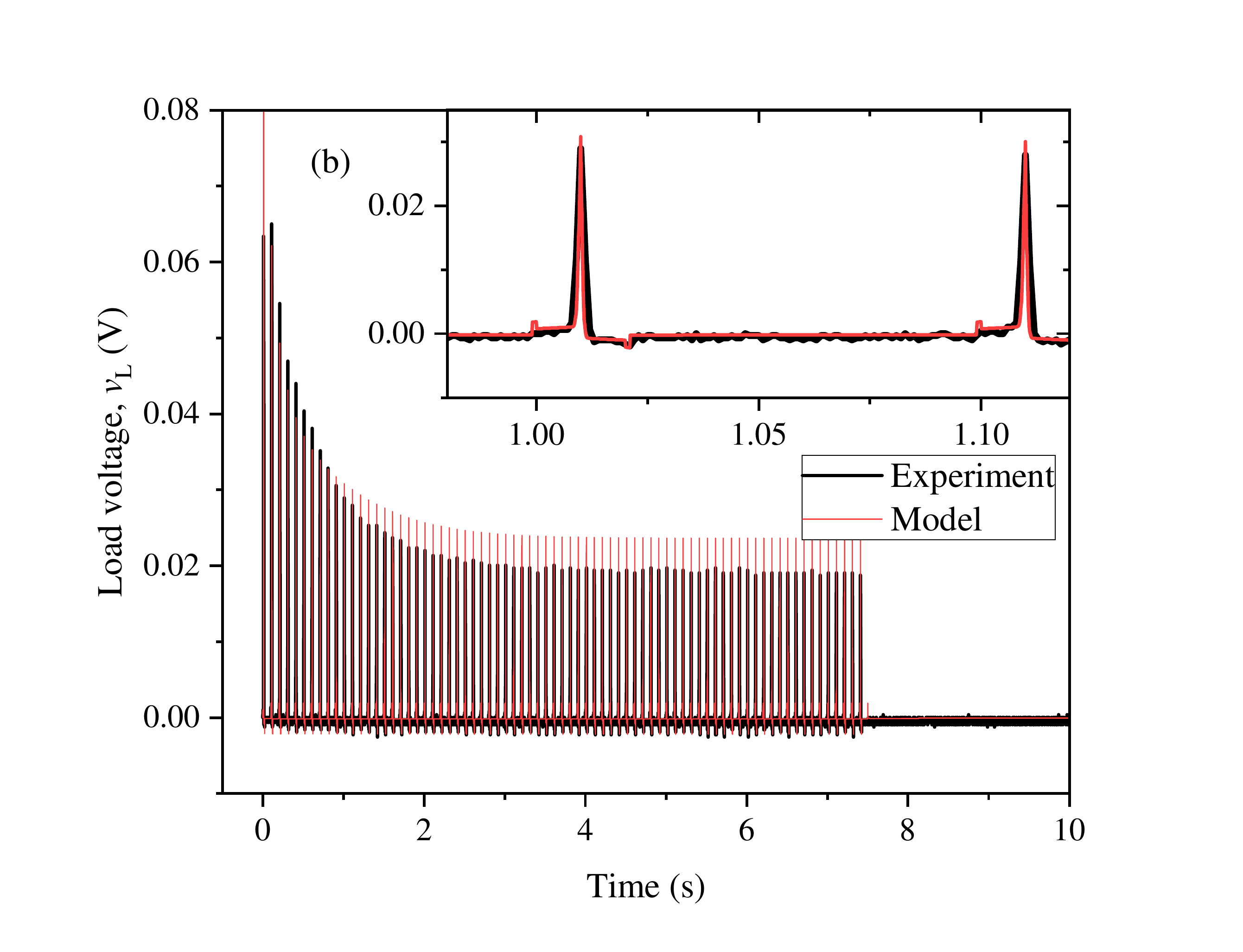}
	\caption{\label{fig:3} Measured (black) and modelled (red) results for primary waveform with 19.8~A maximum, -4.45~A minimum at 10~Hz. (a) The current through the load coil, $i_l$. (b) The voltage across the load coil, $v_l$.   }
\end{figure}

We start by presenting the results of experiments conducted under an input current waveform $i_1\{19.8,-4.45\}$. The current developed through the secondary by $i_{1,max}$ constitutes the `charging phase’ of a current cycle as it is designed to drive current into the load rather than the bridge. The remainder of the cycle we call the `maintenance phase’ as it is designed to compensate the flux accumulated in the transformer core during the charging phase. Simplified diagrams of the charging and maintenance phases are shown in Fig.~\ref{fig:2diagram}.  The current through the load coil as a function of time, $i_L$, and voltage across the load coil, $v_L$, are shown in Fig.~\ref{fig:3}(a) and (b) respectively as black lines. We estimate a 5\% uncertainty in the accuracy of the measured load coil current. Also shown in the figure are the results from the effective circuit modelling, described above, as red lines. 
A full list of parameters used in the model can be found in Appendix~\ref{tab:1}, along with a estimates of the range of their values in the experiment.

Figure~\ref{fig:3}(a) shows that approximately 320~A was supplied to the load coil in under 3~s. This current was induced by a peak voltage across the load coil up to 65~mV during a given cycle (a 20~mV peak once $i_l$ had reached its maximum) that is not fully compensated during the negative part of the input current waveform, Fig.~\ref{fig:3}(b). The figure also shows that our effective-circuit model reproduces; (i) flux-pumping via self-rectification, (ii) the approximate maximum load-coil current and the magnitude of the voltage across the coil and (iii) the approximate waveform of the load-coil current and voltage. 

We note that there are several parameters within the model whose experimental value is not well known and were tuned to generate the agreement shown above. These include the series resistance, $R_{series}$, which in this particular FP design has the same physical origin as $R_{secondary}$, as well as the $n$-value and $I_c$ of the bridge (30 and 462~A respectively), for which we only know their nominal values (35 and 500~A respectively). The actual values may differ from the nominal ones due to tape inhomogeneities and/or mechanical strain during construction and thermal cycling. The resistance of the bridge and load-coil solder joints (450~n$\Omega$) were estimated by matching the decay of current in the load coil once the primary current was turned off, see the data after 10~s in Fig.~\ref{fig:3}(a). For this estimate, we were guided by the $2.5$~$\mu$H inductance of the load coil that was measured at room temperature. 
The behaviour of such FPs relies on the non-linearity of the resistance of the superconductor, equation~\ref{eq:1}. As such, the modelled results are also sensitive to uncertainties accuracy of the measured input current. Other model circuit parameters have similar uncertainties although it was found that the modelled load-coil current maxima and voltage was not as sensitive to them.

\begin{figure}
\centering
	\includegraphics[width=8 cm]{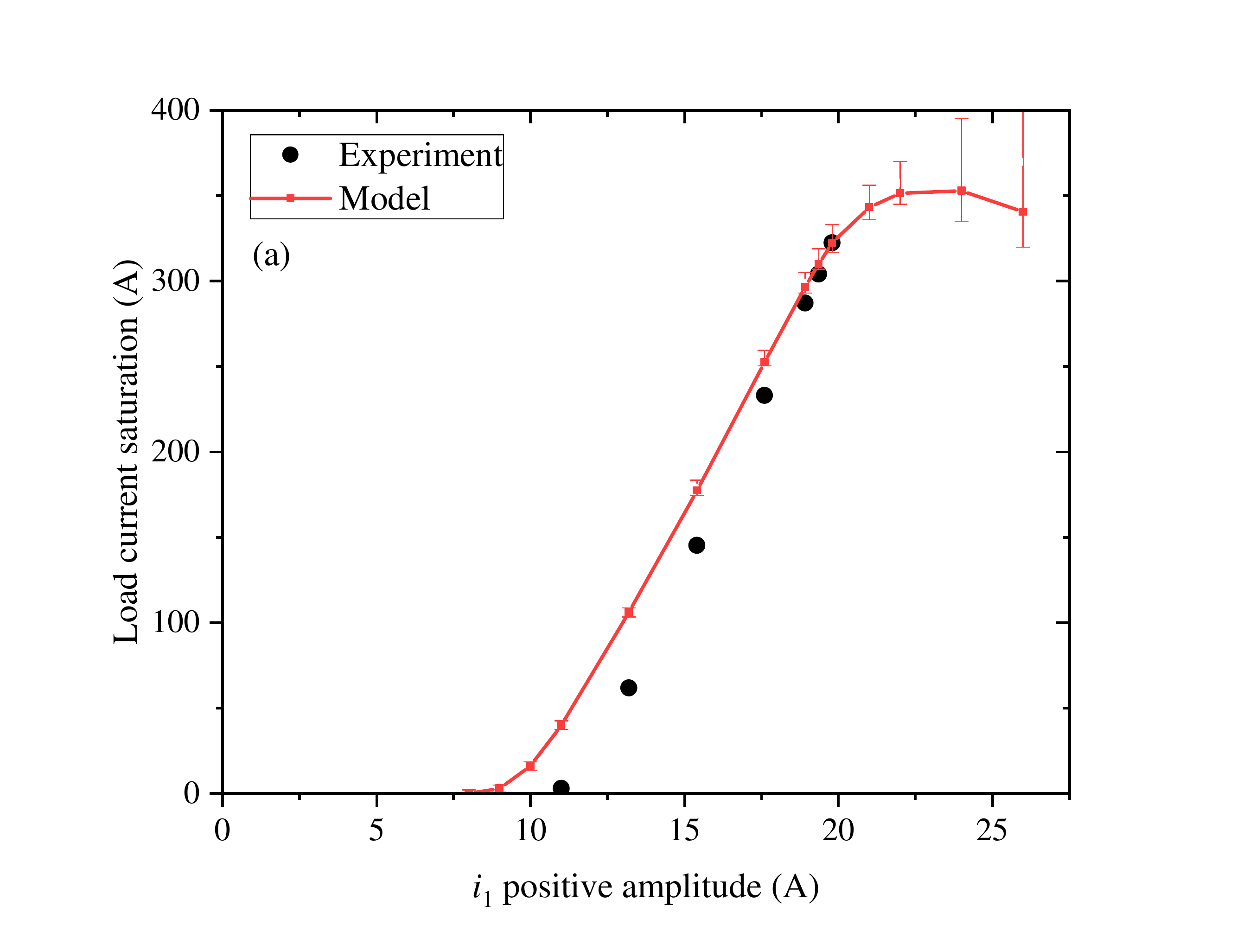}
	\includegraphics[width=8 cm]{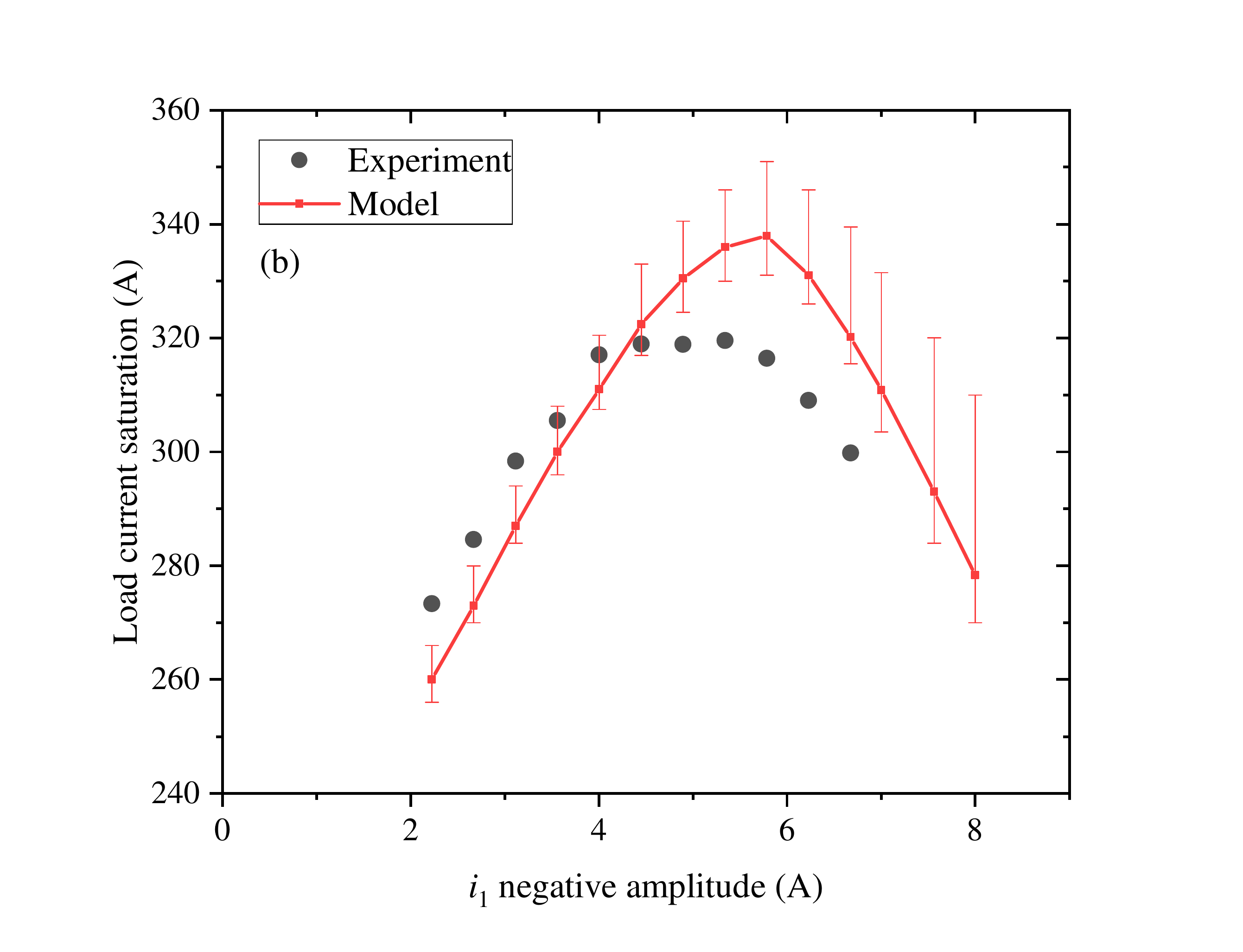}
	\caption{\label{fig:4} Systematic experimental (black symbols) and modelling results (red line and symbols) of the maximum load-coil current. (a) the positive amplitude of $i_1$ is varied for $i_{1,min}=-4.45$~A.  (b) the negative amplitude of $i_1$ is varied for $i_{1,max}=19.8$~A. See Fig.~\ref{fig:1}(b) for example input waveforms. For the modelling results, the data points represent the average load-coil current at its maximum and the error bars represent the range of its current ripple. }
\end{figure}

Confidence in the descriptive ability of the model can however be taken from its ability to capture the dependence of the maximum load coil current on the input waveform. These results are shown in Fig.~\ref{fig:4}, for which panel (a) shows how the maximum load coil current varies with $i_{1,max}$ for $i_{1,min}=-4.45$~A, and panel (b) shows its dependence on $i_{1,min}$ with $i_{1,max}=19.8$~A. 
The error bars for the modelling results represent the range of the current ripple at saturation, whilst the points represent the average value of the load coil current. 
In the effective-circuit model’s near quantitative reproduction of these data sets, only the input waveform was varied in the model to values of $i_1$ measured in the experiment.

\section{Discussion}
We start by discussing insights from our model regarding the achievable load-coil currents, as these appear to reliably reproduce the experimental results, before addressing some potential reasons for the discrepancies between the model predictions and experimental data seen in Fig.~\ref{fig:3} and \ref{fig:4}(b).

Firstly, the modelling results suggest that the maximum load coil current that we achieved experimentally, shown in Fig.~\ref{fig:4}(a), was not limited by the $I_c$ of the load-coil. The model instead shows that with an arbitrarily high coil $I_c$, the load current saturates at the same 320~A for $i_1\{19.8,-4.45 \}$. 
When the coil $I_c$ is the limiting factor within our model, the averaged load current saturates approximately 20~A below the coil $I_c$.

\begin{figure}[htb]
\centering
	\includegraphics[width=0.65\linewidth]{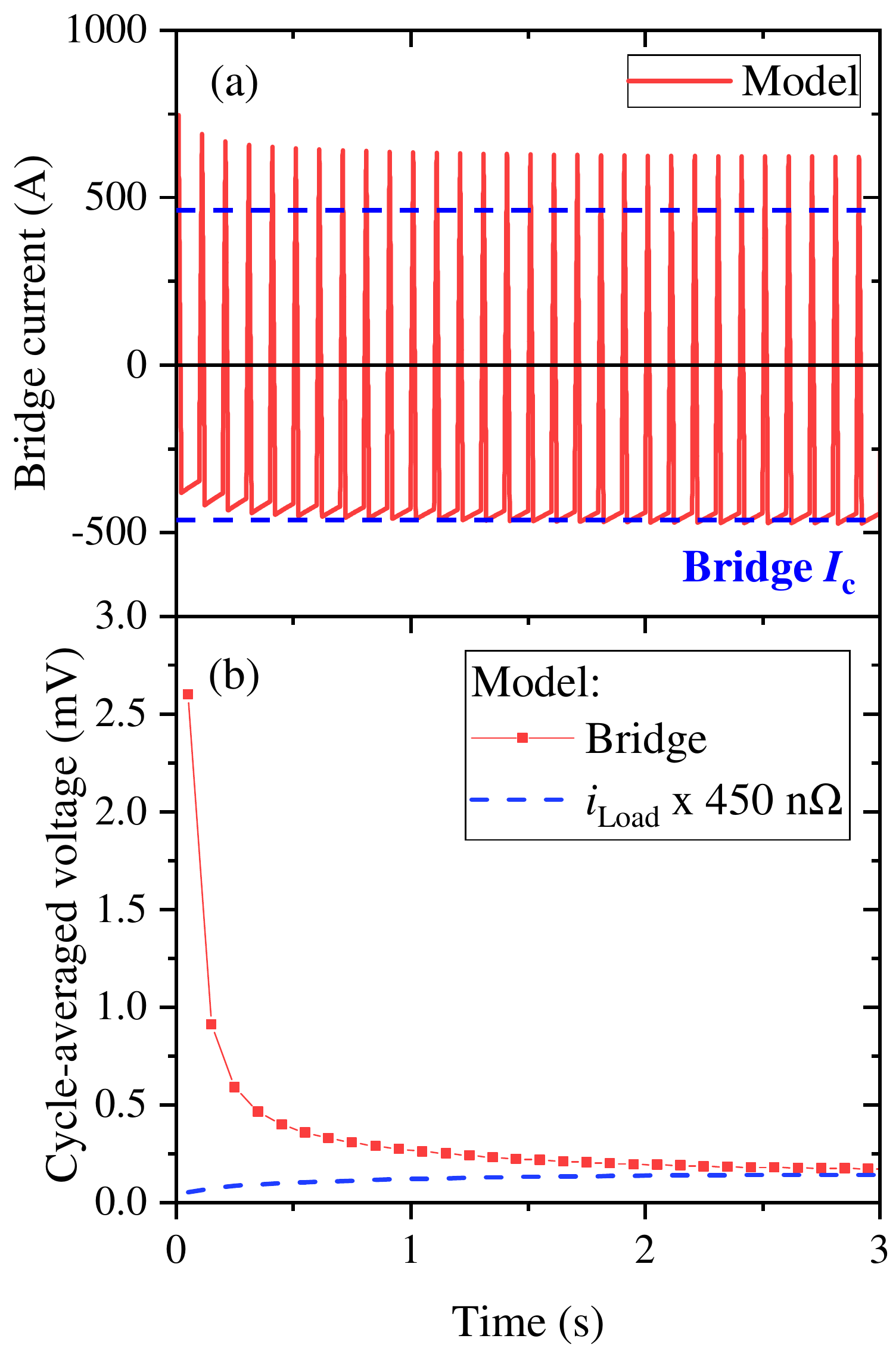}
	\caption{\label{fig:5} Modelling results of the (a) current across the bridge, with the magnitude of the critical current of the bridge, $I_{c,b}$, indicated by dotted lines and (b) average voltage per cycle across the load for the experimental conditions presented in Fig.~\ref{fig:3}, where $i_1$ with a 19.8~A maximum, $-4.45$~A minimum. The dotted blue line indicates the product of the cycle-averaged load current and the load-loop resistance of 450~n$\Omega$.  }
\end{figure}

Rather, it appears to be the net dc voltage developed across bridge that limits our performance. This can be seen in Fig.~\ref{fig:5}, which shows in panel (a) the current across the bridge. A significant negative offset in the bridge current develops over the first second. This has two effects; firstly, the degree to which the bridge current exceeds the critical current of the bridge, $i_{b,max}/I_{c,b}$, is reduced which in turn reduces the voltage developed across the bridge during the charging phase. Secondly, $i_{b,min}/I_{c,b}$ is increased, increasing the negative voltage developed across the bridge during the longer maintenance phase. Panel (b) shows the net-voltage per cycle across the bridge, which quantifies the two effects above. Once the net voltage across the bridge reaches $v_{l} = i_{l}(R_{l}+R_{b})$, the load-coil current has saturated. Here, $R_l+R_b$ is the `load loop' resistance, modelled as 450~n$\Omega$, that accounts for joint-resistances and flux creep. 

The dependence of the maximum load current on $i_{1,max}$ is then straightforward to understand, as $i_{1,max}$ drives current through the bridge in the charging phase to exceed the $I_c$ of the bridge, generating voltage that charges load. For $i_{1,max}>20$~A, the model shows there is no longer a significant increase in the maximum load-coil current. This is due to a voltage developed across the load longer maintenance phase where $i_{b,min}/I_{c,bridge} \approx 1$, and which compensates the voltage developed across the shorter charging phase of the waveform.

The description above aligns with that previously discussed in the literature \cite{geng_hts_2016, geng_maximising_2020, zhai_performance_2021, li_hts_2020}, however it is important to note that there are two potential contributions to the offset in the bridge current. The first is the load current as previously discussed in the literature, and there is a second contribution from an offset-current in the secondary loop. This second component, discussed below, can be manipulated in self-rectifying flux pumps through modification of the $i_1$ (or $v_1$) input. 

The dependence on the $i_{1,min}$ shown in Fig.~\ref{fig:4}(b) displays an optimum. For larger $i_{1,min}$, the decrease in the maximum load-coil current can be understood along similar lines to the situation described above and shown in Fig.~\ref{fig:5}; namely the voltage developed across the bridge during the maintenance phase increases, reducing the net voltage across the bridge. 

\begin{figure*}[htb]
\centering
	\includegraphics[width=13 cm]{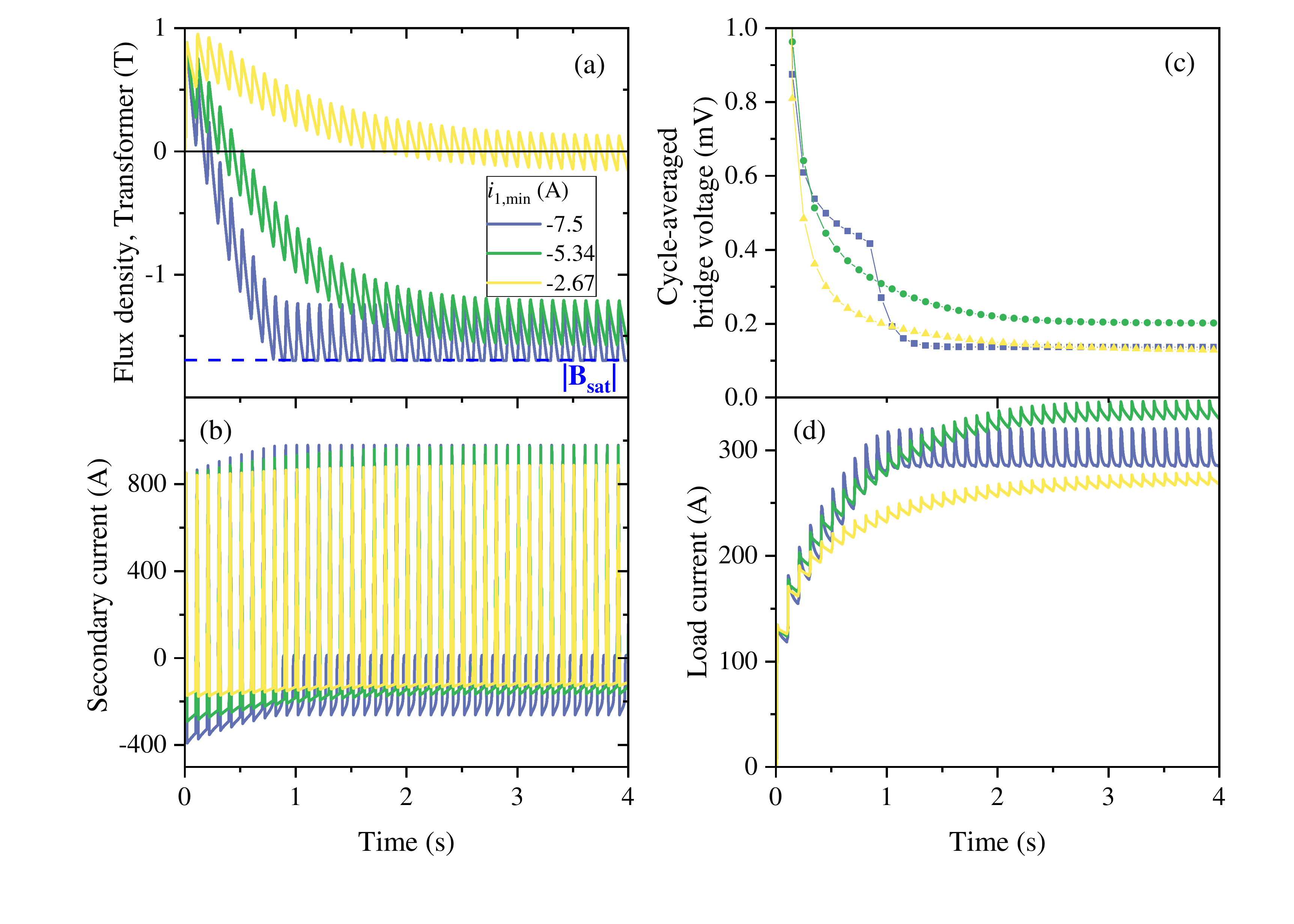}
	\caption{\label{fig:6}  Understanding the maximum $i_l$ as $i_{1,min}$ is varied. Modelling results of (a) flux density in the transformer core, $B $, with the magnitude of the saturation flux density of the core, $B_{\textrm{sat}}$, indicated by a dotted line (b) secondary current $i_2$ (c) the average load voltage per cycle and (d) the load current, $i_l$. }
\end{figure*}

The more interesting case is the reduction of the maximum load coil current at smaller negative current maxima. According to our model, this is not due to transformer saturation, but rather relates to the net flux that develops in the transformer core over multiple cycles, as shown in Fig.~\ref{fig:6}(a). A net flux arises due to a non-zero integral of the input current waveform (and small resistance of the secondary loop) \cite{melkebeek_transformers_2018}. For $i_1\{19.8,-5.34\}$ the integral is $-0.227$ Ampere seconds (A.s) per cycle, compared with $-0.014$~A.s for $i_1\{19.8,-2.67\}$, the yellow curve in Figure with the smaller $i_{1,min}$.  This net flux corresponds to a net dc `offset' current in the secondary loop, as seen in Fig.~\ref{fig:6}(b). That dc offset in the secondary current in turn directly affects the amount by which the bridge $I_c$ is exceeded in the charging phase and the voltage across the bridge, as shown in Fig.~\ref{fig:6}(c). For $i_1\{19.8,-2.67\}$ (yellow curve) that positive dc offset is markedly less than for $i_1\{19.8,-5.34\}$ (green curve) and leads to the reduced voltage across the load and lower maximum load current, Fig.~\ref{fig:6}(d). Using a net offset of flux in the transformer core to improve the performance of a self-rectifying flux pump has been qualitatively described earlier \cite{geng_maximising_2020}, and this work now shows a quantitative description of this strategy.

\begin{figure}[htb]
\centering
	\includegraphics[width=8 cm]{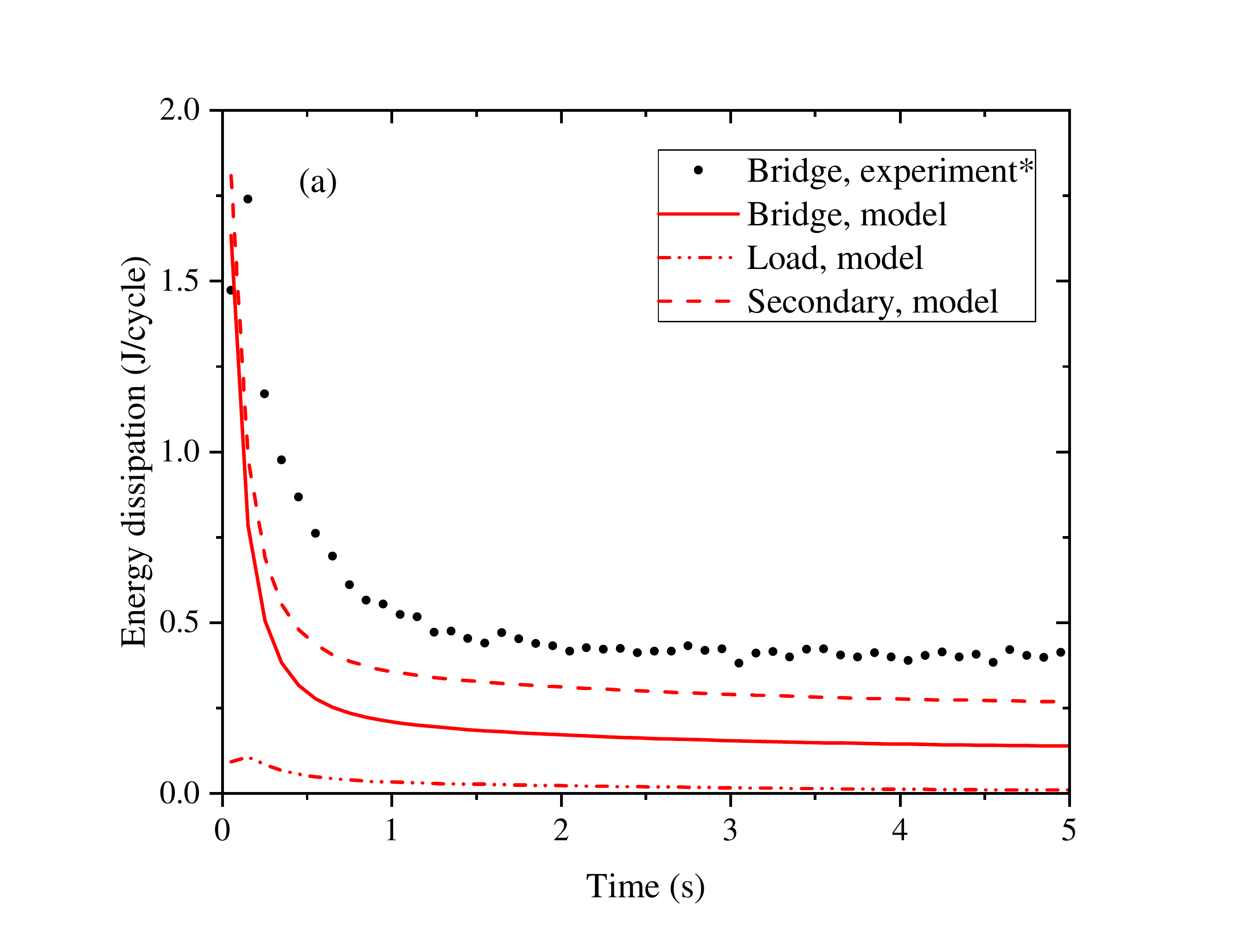}
	\includegraphics[width=6 cm]{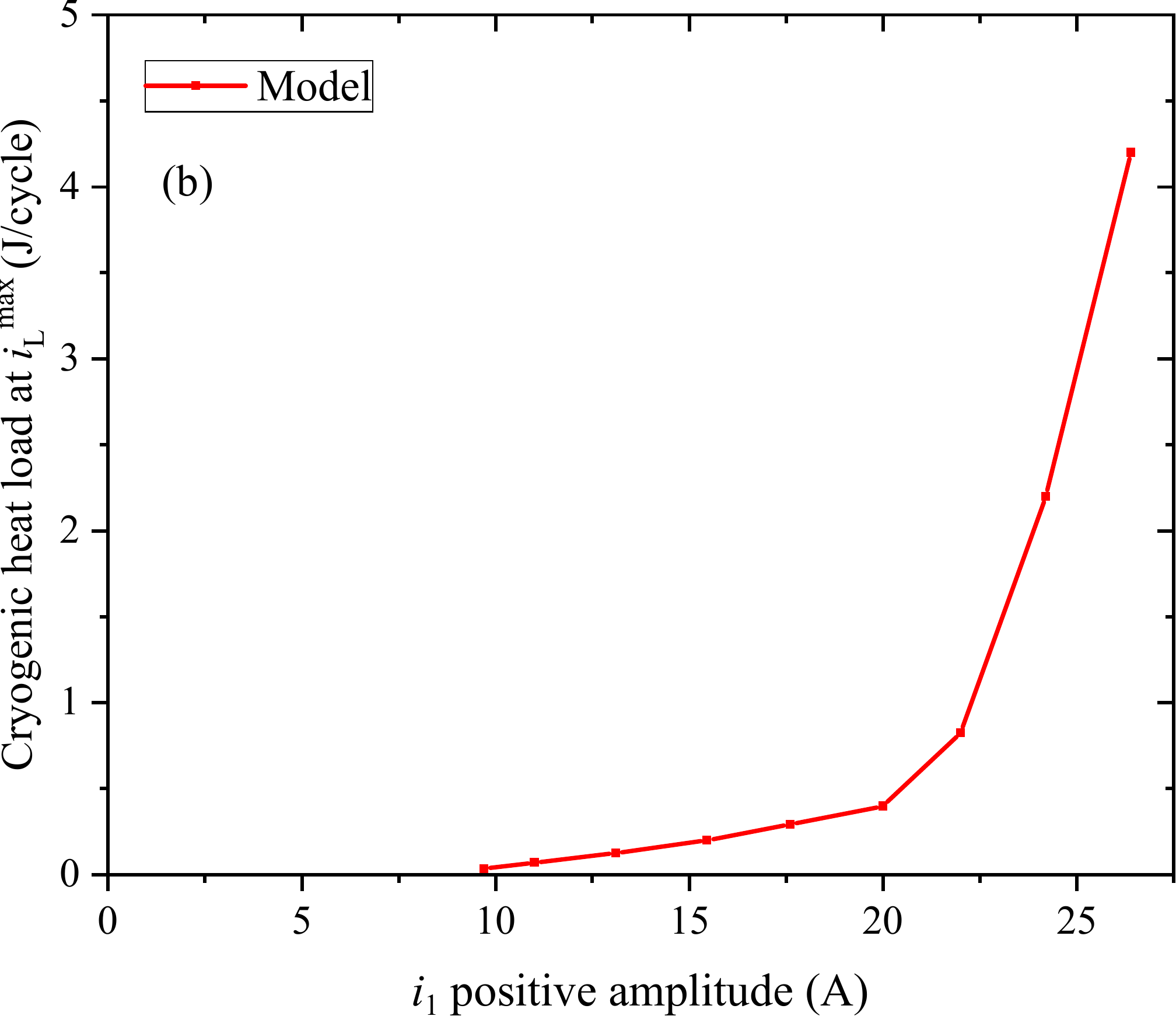}
	\caption{\label{fig:7} Energy dissipation across various device elements under $i_1\{19.8,-4.45\}$ current input; black symbols for the bridge as estimated from experimental data (see text), red lines show results from the model. (b) Total heat dissipation into the cryogenic environment as a function of $i_{1,max}$ with $i_{1,min}=-4.45$~A once the maximum $i_l$ has been reached ($t=5$~s) – modelling results only.  }
\end{figure}

A strategy of increasing the $i_1$ current amplitude to increase the maximum load-coil current may not be desirable. For example, the heat dissipated across various parts of the circuit rapidly increases, as shown in Fig.\ref{fig:7}. Figure~\ref{fig:7}(a) shows the heat dissipated per cycle across the bridge as estimated from experimental results. This is regarded as an estimate because it was not possible to directly measure the experimental current through either element. Instead, determination of the bridge current required the approximation that the current across the secondary is $N$ times that of the primary. This approximation is imperfect, as we do not have an ideal transformer. Also shown in the panel is the heat dissipated by the various elements on the secondary side of the transformer, the `cryogenic side’ (see Fig.~\ref{fig:2}), as calculated within our model. Figure~\ref{fig:7}(b) shows the degree to which the total heat dissipated on the cryogenic side, calculated after the load current has saturated at $t=5$~s, increases with input current maxima. 

We have shown that the model behaviour and experimental results show broad agreement, and now we discuss their discrepancies. Most noticeable is the ramp-rate of the dc component of the load-coil current, in particular the very first cycle of operation, as highlighted in the inset to Fig.~\ref{fig:3}(a). There is also some quantitative inaccuracy in the dependence of $i_{l,max}$ on $i_{1,min}$ predicted by the model, as shown in Fig.~\ref{fig:4}(b). 

We identify three general potential reasons for these discrepancies. Firstly, we may have model parameter values that inadequately represent the experiment. This is likely to be the case to some degree. However after investigating several different parameter configurations that reproduce the experimental data sets, we find that the same physical interpretations from the model that are discussed above. Secondly, our electromagnetic effective-circuit model may be missing important circuit elements. We do not believe this to be the case. 
Thirdly, our effective circuit model may be missing important physics. Perhaps the most salient effects not accounted for within our effective circuit model are thermal effects (for example, heating of superconducting components due to the energy dissipation highlighted in Fig.~\ref{fig:7}) and ac loss or hysteresis in the transformer core \cite{melkebeek_transformers_2018}. In this regard, it is noteworthy that the heat dissipation across the bridge estimated from experiments is approximately twice that calculated by the model, Fig.~\ref{fig:7}(a).
We expect loss relating to the current ramp rate, $di/dt$, in the superconducting components to be relatively minor \cite{song_voltage-current_2018}, however we note that we have large $di/dt$ values of up to approximately $9\times10^4$~A.s$^{-1}$ and our asymmetric waveform presents additional complications over the standard analysis presented in literature, see e.g. \cite{norris_calculation_1970, chen_scaling_2015}. We also note that the degree to which these effects are significant will depend on the details of the experimental set up. Future experiments and extensions to the current effective circuit model will look to quantitatively incorporate such physics, as has recently been done in other areas of superconductor power electronics \cite{trillaud_essential_2021}, and assess the degree to which they improve agreement with experiment.

\section{Summary}
In summary, we demonstrated the operation of an ultra-compact self-rectifier flux pump that supplied up to 320~A dc to a superconducting coil and a peak output voltage of up to 60~mV. Whilst that is not the largest current or voltage demonstrated from a self-rectifier to date, our device is exceptionally compact, simple in architecture and demonstrated that the secondary winding need not be made from a superconductor. These features, along with their minimal heat-leak into the cryogenic environment from thermal conduction, highlight how attractive this class of flux pump is for applications with demanding size, weight and power limitations. We also developed an electromagnetic effective-circuit model of the flux pump, encompassing the transformer and all superconducting components, in order to understand and predict the behaviour of this class of flux pump. A comparison of its predictions to the experimental results showed that many key aspects of the experiment are captured by this model, such as the maximum load current and its dependence on the input current to the transformer. Work is currently ongoing to understand why certain aspects of the experiment were not accurately reproduced by the model, such as the load current ramp rate, with this work including investigations into the effect of including additional physics into the model. Nevertheless, the predictive successes of our model and its inclusion of all key components of the flux pump system, show it to be a powerful tool in the design and optimization for applications of this class of flux pump.  

\section{Acknowledgements}
This work was supported in part by New Zealand Ministry of Business, Innovation and Employment (MBIE) by through the project `High Magnetic Field Electric Propulsion for Space', contract number RTVU2003 and
the Strategic Science Investment Fund `Advanced Energy Technology Platforms' under Contract RTVU2004.

\section{Appendix}
In the table below we present a comparison of model parameters used in this work and experimental estimates where available.

\begin{table*}[htb]
    \centering
    \begin{tabular}{ l |  l  |  l  |  l }
         Circuit element &  Parameter [units] &  Experimental range &  Value in model \\  \hline\hline
         Bridge & $I_c$ [A] & 470-550                   & 462 \\ 
                & $n$       & 25-35                     & 30 \\ 
                & $l$ [m]   & 0.09-0.11                 & 0.1  \\ 
                & $\tau$ [$\mu$m]   & 18-22     & 20  \\ 
                & $L$ [H]   & (10-200)$ \times10^{-9}$  & $10 \times10^{-9}$ \\ \hline
         Load   & $I_c$ [A] & 400-550                   & 360 \\ 
                & $n$       & 15-35                     & 20 \\ 
                & $l$ [m]   & 1.85-1.95                 & 1.9  \\
                & $\tau$ [$\mu$m]& 18-22        & 20 \\ 
                & $L$ [$\mu$H]   & 2-3         & 2.0  \\\hline
       Secondary& $N$       & 1                   & 1 \\  
                & $l$ [m]   & 0.08-0.12           & 0.1  \\ 
                & $A$ [mm$^2$]   & 32-40        & 36 \\
                & $\rho_{\textrm{Cu}}$ [n$\Omega$.m]   & 1.9-4.0  & 3.8 \\\hline
        Primary & $N$       & 45                  & 45 \\  
                & $R$ [$\Omega$]   & 0.55 -0.6          & 0.57  \\ \hline
        Resistances & $R_{\textrm{series}}$ [n$\Omega$]   &     -    & 100 \\
                    & $R_{\textrm{bridge}}$ [n$\Omega$]   &  150-750 & 400  \\
                    & $R_{\textrm{load}}$ [n$\Omega$]   &    50-600  & 50  \\ \hline
        Core    & $A$ [mm$^2$]   & 350-450        & 400 \\
                & $l$ [mm]         & 90-110         & 100 \\
                & $B_{\textrm{sat}}$ [T] & 1.5-2.1  & 1.7 \\
                & $H_{\textrm{sat}}$ [A.m${-1}$] & 200-400  & 300 \\
                & Leakage reluctance [H$^{-1}$] &  (0.01-1)$ \times10^{9}$ & $0.1 \times10^{9}$ \\
                & Gap [mm] &  0.05-0.5 & 0.1 \\
    \end{tabular}
    \caption{\label{tab:1}A comparison of model input parameters and estimates for the range of experimental values of those parameters.}
\end{table*}


\end{document}